\documentclass[10pt,conference]{IEEEtran}
\AtBeginDocument{%
  }

\usepackage{setspace}
\usepackage{bm}
\usepackage{xcolor}
\usepackage{booktabs}
\usepackage{microtype}
\usepackage{amsfonts}
\usepackage{xspace}
\usepackage{etoolbox}
\usepackage{array}
\usepackage{microtype}
\usepackage{multirow}
\usepackage{framed}
\usepackage{caption}
\usepackage{threeparttable}
\usepackage{fontawesome}
\usepackage[labelformat=simple]{subcaption}
\usepackage[ruled,vlined]{algorithm2e}
\usepackage{cite}
\usepackage{url}

\usepackage{footnote}
\makesavenoteenv{tabular}

\SetCommentSty{mycommfont}

\SetKwInOut{Input}{Input}
\SetKwInOut{Output}{Output}
\SetKwInOut{Except}{Except.}
\SetKw{None}{nil}
\SetKw{Cont}{continue}
\SetKw{Break}{break}
\SetKwProg{Fn}{Function}{:}{}
\SetKwFunction{Initp}{InitPattern}
\SetKwFunction{Nextp}{NextPattern}
\SetKwFunction{Initss}{InitSerSched}
\SetKwFunction{Nextss}{NextSerSched}
\SetAlCapFnt{\small}
\SetAlCapNameFnt{\small}
\LinesNumbered
\DontPrintSemicolon

\usepackage{tikz}
\usetikzlibrary{positioning}
\tikzstyle{period} = [draw=white, fill=gray!30, thick,
    rectangle, rounded corners, minimum size=9ex]
\usepackage{subcaption}
\usepackage{soul}
\newcommand{\encircled}[1]{%
    \tikz[baseline=(char.base)]{
        \node[shape=circle, draw, fill=black, text=white, font=\scriptsize, inner sep=0.5pt] (char) {#1};}
}

\AtBeginEnvironment{verbatim}{\setlist[trivlist]{nosep}}
\usepackage{flushend}
\usepackage{balance}
\usepackage{graphicx}
\usepackage{amsthm}
\usepackage{amsmath}
\usepackage{amssymb}

\usepackage{bbding}
\usepackage{array}
\usepackage{multirow}
\usepackage{listings}
\usepackage{color}
\usepackage{tcolorbox}
\usepackage{xcolor,colortbl}
\usepackage{enumitem}
\usepackage{adjustbox}
\usepackage[normalem]{ulem}
\usepackage{pifont}
\usepackage{wasysym}

\usepackage[hang,flushmargin]{footmisc}

\definecolor{nord0}{HTML}{2E3440}
\definecolor{nord1}{HTML}{3B4252}
\definecolor{nord2}{HTML}{434C5E}
\definecolor{nord3}{HTML}{4C566A}
\definecolor{nord4}{HTML}{D8DEE9}
\definecolor{nord5}{HTML}{E5E9F0}
\definecolor{nord6}{HTML}{ECEFF4}
\definecolor{nord7}{HTML}{8FBCBB}
\definecolor{nord8}{HTML}{88C0D0}
\definecolor{nord9}{HTML}{81A1C1}
\definecolor{nord10}{HTML}{5E81AC}
\definecolor{nord11}{HTML}{BF616A}
\definecolor{nord12}{HTML}{D08770}
\definecolor{nord13}{HTML}{EBCB8B}
\definecolor{nord14}{HTML}{A3BE8C}
\definecolor{nord15}{HTML}{B48EAD}
\definecolor{seen}{HTML}{C8D2EB}
\definecolor{unseen}{HTML}{FDF0CE}

\definecolor{downcolor}{HTML}{5573A5}
\definecolor{upcolor}{HTML}{B85B59}

\usetikzlibrary{shapes,positioning}
\makeatletter
\let\old@lstKV@SwitchCases\lstKV@SwitchCases
\def\lstKV@SwitchCases#1#2#3{}
\makeatother
\usepackage{lstlinebgrd}
\makeatletter
\let\lstKV@SwitchCases\old@lstKV@SwitchCases

\lst@Key{numbers}{none}{%
    \def\lst@PlaceNumber{\lst@linebgrd}%
    \lstKV@SwitchCases{#1}%
    {none:\\%
        left:\def\lst@PlaceNumber{\llap{\normalfont
                \lst@numberstyle{\thelstnumber}\kern\lst@numbersep}\lst@linebgrd}\\%
        right:\def\lst@PlaceNumber{\rlap{\normalfont
                \kern\linewidth \kern\lst@numbersep
                \lst@numberstyle{\thelstnumber}}\lst@linebgrd}%
    }{\PackageError{Listings}{Numbers #1 unknown}\@ehc}}
\makeatother

\definecolor{rowcolor}{HTML}{ECEFF4}

\setlist[itemize]{leftmargin=*}

\definecolor{dkgreen}{rgb}{0,0.6,0}
\definecolor{gray}{rgb}{0.5,0.5,0.5}
\definecolor{mauve}{rgb}{0.58,0,0.82}

\definecolor{bg}{HTML}{F8F9FB}  % FAFAFA
\definecolor{bgc}{HTML}{FCF6E4}
\usepackage[framemethod=TikZ]{mdframed}

\mdfdefinestyle{MyFrame}{%
    linecolor=black,
    outerlinewidth=0.3pt,
    roundcorner=5pt,
    skipabove = 5.5pt,
    skipbelow = 5.5pt,
    innertopmargin=5.5pt, %\baselineskip,
    innerbottommargin=5.5pt, %\baselineskip,\baselineskip,
    innerrightmargin=5.5pt,
    innerleftmargin=5.5pt,
    backgroundcolor=bg, %gray!25!white
    }

\newcommand{\etal}{\hbox{\emph{et al.}}\xspace}
\newcommand{\eg}{\hbox{\emph{e.g.}}\xspace}
\newcommand{\ie}{\hbox{\emph{i.e.}}\xspace}
\newcommand{\etc}{\hbox{\emph{etc.}}\xspace}

\newcommand{\name}{\textsc{CodeCleaner}\xspace}

\setlist[itemize]{leftmargin=*}
\setlist[enumerate]{leftmargin=*}
\newlist{steps}{enumerate}{1}
\setlist[steps, 1]{label = \textbf{RQ\arabic*.}}

\begin{document}

\title{\name: Elevating Standards with A Robust Data Contamination Mitigation Toolkit
}

\author{
\IEEEauthorblockN{Jialun Cao}
\IEEEauthorblockA{The Hong Kong University of Science \\and Technology\\
Hong Kong, China}
\and
\IEEEauthorblockN{Songqiang Chen}
\IEEEauthorblockA{The Hong Kong University of Science \\and Technology\\
Hong Kong, China}
\and
\IEEEauthorblockN{Wuqi Zhang}
\IEEEauthorblockA{The Hong Kong University of Science \\and Technology\\
Hong Kong, China}
\and
\IEEEauthorblockN{Hau Ching Lo}
\IEEEauthorblockA{The Hong Kong University of Science \\and Technology\\
Hong Kong, China}
\and
\IEEEauthorblockN{Yeting Li}
\IEEEauthorblockA{Institute of Software\\
Chinese Academy of Sciences\\
University of Chinese Academy of Sciences}
\and
\IEEEauthorblockN{Shing-Chi Cheung}
\IEEEauthorblockA{The Hong Kong University of Science \\and Technology\\
Hong Kong, China}
}

% \begin{CCSXML}
% <ccs2012>
%    <concept>
%        <concept_desc>Software and its engineering~Software testing and debugging</concept_desc>
%        <concept_significance>500</concept_significance>
%        </concept>
%  </ccs2012>
% \end{CCSXML}
% \begin{CCSXML}
% <ccs2012>
%    <concept>
%        <concept_id>10011007.10011074.10011092.10011782</concept_id>
%        <concept_desc>Software and its engineering~Automatic programming</concept_desc>
%        <concept_significance>500</concept_significance>
%        </concept>
%  </ccs2012>
% \end{CCSXML}

% \ccsdesc[500]{Software and its engineering~Automatic programming}
% \ccsdesc[500]{Software and its engineering~Software testing and debugging}

\maketitle

%
% This command processes the author and affiliation and title information and builds
% the first part of the formatted document.
%\IEEEpeerreviewmaketitle
% \settopmatter{printfolios=true}

%\hspace*{\fill} \\
%\hspace*{\fill} \\
%\hspace*{\fill} \\
%{\color{gray}This is the author's version of the work. It is posted here for your personal use. Not for redistribution. The definitive version was published in the proceedings of XXX.}

%\newpage

\begin{abstract}
Data contamination presents a critical barrier preventing widespread industrial adoption of advanced software engineering techniques that leverage code language models (CLMs). This phenomenon occurs when evaluation data inadvertently overlaps with the public code repositories used to train CLMs, severely undermining the credibility of performance evaluations. For software companies considering the integration of CLM-based techniques into their development pipeline, this uncertainty about true performance metrics poses an unacceptable business risk.
Code refactoring, which comprises code restructuring and variable renaming, has emerged as a promising measure to mitigate data contamination. It provides a practical alternative to the resource-intensive process of building contamination-free evaluation datasets, which would require companies to collect, clean, and label code created after the CLMs' training cutoff dates. However, the lack of automated code refactoring tools and scientifically validated refactoring techniques has hampered widespread industrial implementation. 

To bridge the gap, this paper presents the first systematic study to examine \textit{the efficacy of code refactoring operators at multiple scales} (method-level, class-level, and cross-class level) and \textit{in different programming languages}. In particular, we develop an open-sourced toolkit, \name, which includes 11 operators for Python, with nine method-level, one class-level, and one cross-class level operator. 
We elaborate on the rationale for why these operators could work to resolve data contamination and use both data-wise (\eg, N-gram matching overlap ratio) and model-wise metrics (\eg, perplexity) to quantify the efficacy after operators are applied. A drop of 65\% overlap ratio is found when applying all operators in \name, demonstrating their effectiveness in addressing data contamination. 
Additionally, we migrate four operators to Java, showing their generalizability to another language. We make \name online available at \url{https://github.com/ArabelaTso/CodeCleaner-v1/} to facilitate further studies on mitigating CLM data contamination.

\end{abstract}


\section{Introduction}\label{sec:intro}

The software industry has increasingly embraced techniques to leverage code language models (CLMs) as powerful tools for various software development and maintenance tasks~\cite{deng2023large,deng2024large,xia2024fuzz4all,xia2023automated}. However, a critical concern has emerged regarding their evaluation: these models, trained on vast code repositories, may have already encountered the test data during their training phase. This phenomenon, known as \textbf{\textit{data contamination}}~\cite{dataContamination2023,sainz-etal-2023-nlp-contam,taskContamination2023,li2023open,cao2024concerned}, can lead to artificially inflated performance assessment, raising questions about the true capabilities of these techniques. Recent studies~\cite{sainz-etal-2023-nlp-contam,dataContamination2023,taskContamination2023,balloccu-etal-2024-leak,tirumala2022memorization,kandpal2022deduplicating,schick2020s,magar2022data} have highlighted this issue as a major threat to the reliable assessment of CLM-based techniques. The threat poses an unacceptable business risk in adopting these techniques.

\textit{\textbf{Challenge --}}
However, identifying and preventing data contamination is non-trivial. First, in the era of large language models (LLMs), training corpora is usually tremendously massive. As pointed out by a recent survey~\cite{liu2024datasetslargelanguagemodels}, the total data size for pre-training corpora surpasses 774.5 TB, making it difficult to analyze the entire corpora for data contamination~\cite{topkmia2023,extracting21,touvron2023llama}.
In addition, data contamination could be indirectly introduced inadvertently. For example, LLMs may acquire programming languages not only from code hosting platforms such as GitHub and GitLab but also from blogs, open-source forums, or social media feeds. Third, thoroughly eliminating data contamination from CLMs is hard due to the nature of copy-and-paste coding practices~\cite{copypaste}. Software developers reuse, reference, or adapt existing source codes to prevent reinventing the wheels while enhancing development efficiency.
Besides, CLMs' training corpora will increasingly contain AI-written codes with the prevalent use of AI programming assistants such as Github Copilot~\cite{githubcopilot}. As reported by Github on Aug 21, 2024, 
{{nearly all (97\%) developers use AI coding tools}}~\cite{githubAI2024}. In other words, identifying and preventing data contamination is challenging. 

Recent studies have been aware of the need to alleviate the data contamination threat to the validity in evaluating their proposed CLM-based techniques~\cite{generalization2024,wang2024leakdetection,mradopt2024}. 
One solution is to collect data (\eg, code) uploaded after CLMs' release date (\ie, the cut-off date). Yet, as mentioned above, considering the increasing use of AI assistants in software development, collecting new code is not an effective solution. An alternative is to construct benchmarks or adapt them from crawled open-source platforms manually. Representative benchmarks like HumanEval~\cite{humaneval} and SWE-bench~\cite{jimenez2024swebench} were once believed to be data contamination-free. However, as CLMs evolve fast, many prior released benchmarks were reported to appear in recent-released CLMs~\cite{matton2024leakage}, invalidating their effectiveness for addressing the data contamination issue.

\textit{\textbf{Research Gap --}} \textbf{\textit{Code refactoring}}~\cite{shirafuji2023refactoring,VJBench23} (\eg, changing code structures and renaming the variables) is considered a lightweight solution to mitigate data contamination. However, four outstanding challenges remain to be solved for effective mitigation. 
First, \textbf{how may code be refactored \textit{without altering its original semantics}?} Unlike perturbing text and image data, perturbing code while maintaining its syntactic, semantic, and logic requirements is difficult. Arbitrary modifying characters (\eg, adding/deleting/replacing characters) in the code string could easily ruin the code's grammar or functionalities, making it uncompilable or malfunctioning. 
Second, \textbf{it is unclear how various code refactoring operators actually perform in practice}. While previous studies have applied refactoring operators both manually and automatically, they typically lack comparative analysis between pre- and post-refactored code. This absence of direct comparison makes validating whether these operators can reduce data contamination difficult. Recent research has revealed an unexpected outcome: certain refactoring operations may actually increase the model's familiarity with the code, potentially counteracting their intended purpose~\cite{cao2024concerned}.
Third, \textbf{the effectiveness of code refactoring techniques can vary significantly across different programming languages}. What proves successful in one language may not translate effectively to another. This variability is particularly relevant given recent research~\cite{wang2024exploring} demonstrating how language models perform differently across programming languages. These findings underscore the need to better understand both how data contamination manifests differently across programming languages and how refactoring strategies must be adapted accordingly. 
Fourth, \textbf{the scope of code refactoring - ranging from individual methods to entire classes or projects - presents another critical challenge}. While method-level refactoring focuses on maintaining internal coherence within a single method, larger-scale refactoring must address the additional complexity of managing relationships between multiple functions and classes. This fundamental difference suggests that refactoring strategies need to be specifically designed and evaluated for different levels of code granularity.

\begin{figure}[t!]
    \centering
    % \vspace{-0ex}
    \includegraphics[width=1.0\linewidth]{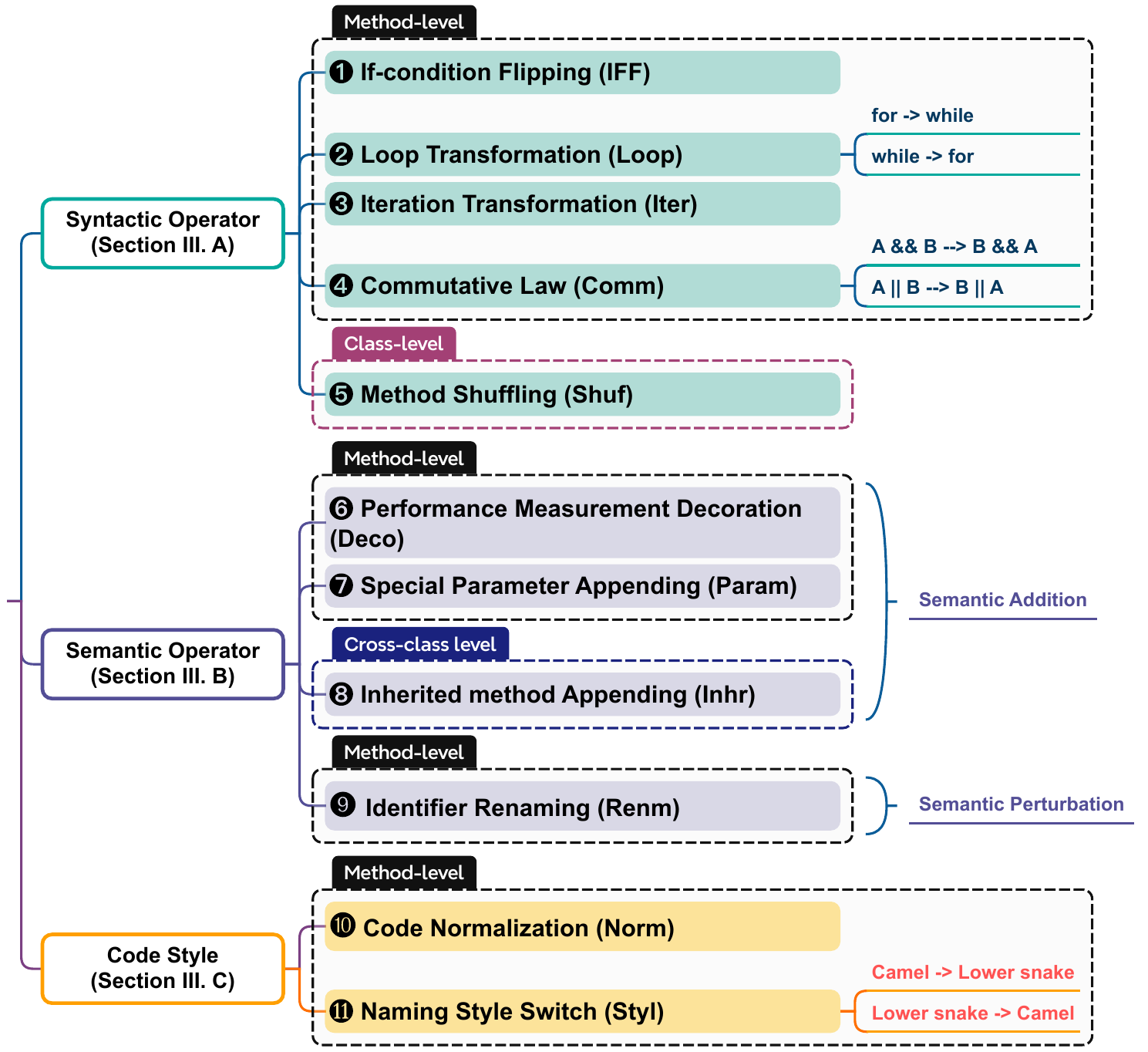}
    \setlength{\abovecaptionskip}{-5pt}
    \setlength{\belowcaptionskip}{-15pt}
    \caption{Code Refactoring Operators in \name}
    \label{fig:opr}
\end{figure} 

To explore these challenges, we implemented eleven code refactoring operators for either method-, class-, or cross-class-levels. These refactoring operators are meticulously handled to alter the code's \textit{syntactic}, \textit{semantics}, and \textit{code style} without affecting the original semantic integrity and maintaining correct syntax, as shown in Figure~\ref{fig:opr}. Upon the implementation, we conducted the first systematic study assessing the efficacy of code refactors to data contamination on CLMs. We focus on \textit{code data} rather than text or image data because of their relevance to software engineering areas. To determine the severity of the contamination, we consider two bunches of measurements, \ie, \textit{perplexity}~\cite{jelinek1977perplexity} and its variants and \textit{n-gram} overlap ratio. In particular, we use \textit{the Stack} as training data corpus and use the contaminated code extracted from \textit{the Stack}~\cite{url_stackv1hf} as a prior work~\cite{cao2024concerned} did to conduct our experiment. We apply different operators to it and investigate whether the perplexity or n-gram overlap ratios show significant changes. 

We study four research questions (RQs) accordingly.

\begin{itemize}
    \item \textbf{RQ1. How do different code refactoring operators affect contamination severity for method-level Python code?} We apply each code refactoring operator to method-level codes and study how the contamination severity changes after each refactoring operator is applied.

    \item \textbf{RQ2. How do class-level and method-level code refactoring operators affect contamination severity in class-level Python code?} When a simple refactoring operator is applied to larger-scale code snippets, the effectiveness may be reduced. We thus investigate the effectiveness of code refactoring operators in class-level Python code.
        
    \item \textbf{RQ3. How does data contamination differ between programming languages?} Most existing works study data contamination in Python, while it is unclear how this issue varies between different programming languages. We thus sample four other programming languages (\ie, Java, C, and Rust) using the same sampling criteria and strategy as a prior work~\cite{cao2024concerned} did for Python and checking the differences.
    
    \item \textbf{RQ4. How do code refactoring operators differ in Python and Java?} We further implement several key code refactoring operators for Java and compare that for Python. The correlations may provide insights into the generalizability of different operators.

\end{itemize}

Our experiment shows the potential of \name. In particular, 
method-level operators can reduce 35.89\% overlap with the training set on average. Semantic operators are generally much more effective than syntactic ones. When all operators are considered, a dramatic drop of 65\% overlap ratio can be achieved, demonstrating operators' exceptional capabilities in addressing data contamination.

\noindent\textbf{{Contributions --}} 
Our contribution is summarized as follows. 
\begin{itemize}
    \item \textbf{Novelty --} 
    We present the first study on the effectiveness of code refactoring operators in mitigating data contamination. It investigates both method-level and class-level refactoring operators and their effectiveness. We also explore the severity of data contamination in different programming languages and the effectiveness of operators in them.
    
    \item \textbf{Usefulness --} We provide the first off-the-shelf automated code refactoring toolkit, \name, to mitigate data contamination in evaluating CLMs. It includes 11 refactoring operators applicable to Python and Java codes, covering structure, semantics, and code style refactoring.

    \item \textbf{Impact --}
    The design of most code refactoring operators is language-agnostic and thus can be transferred to other programming languages. Our findings shed light on further studying the data contamination in CLM-based techniques.
    
\end{itemize}

\section{Motivating Example}\label{sec:moti}

\begin{figure}[th]
    \centering
    % \vspace{-0ex}
    \includegraphics[width=1.0\linewidth]{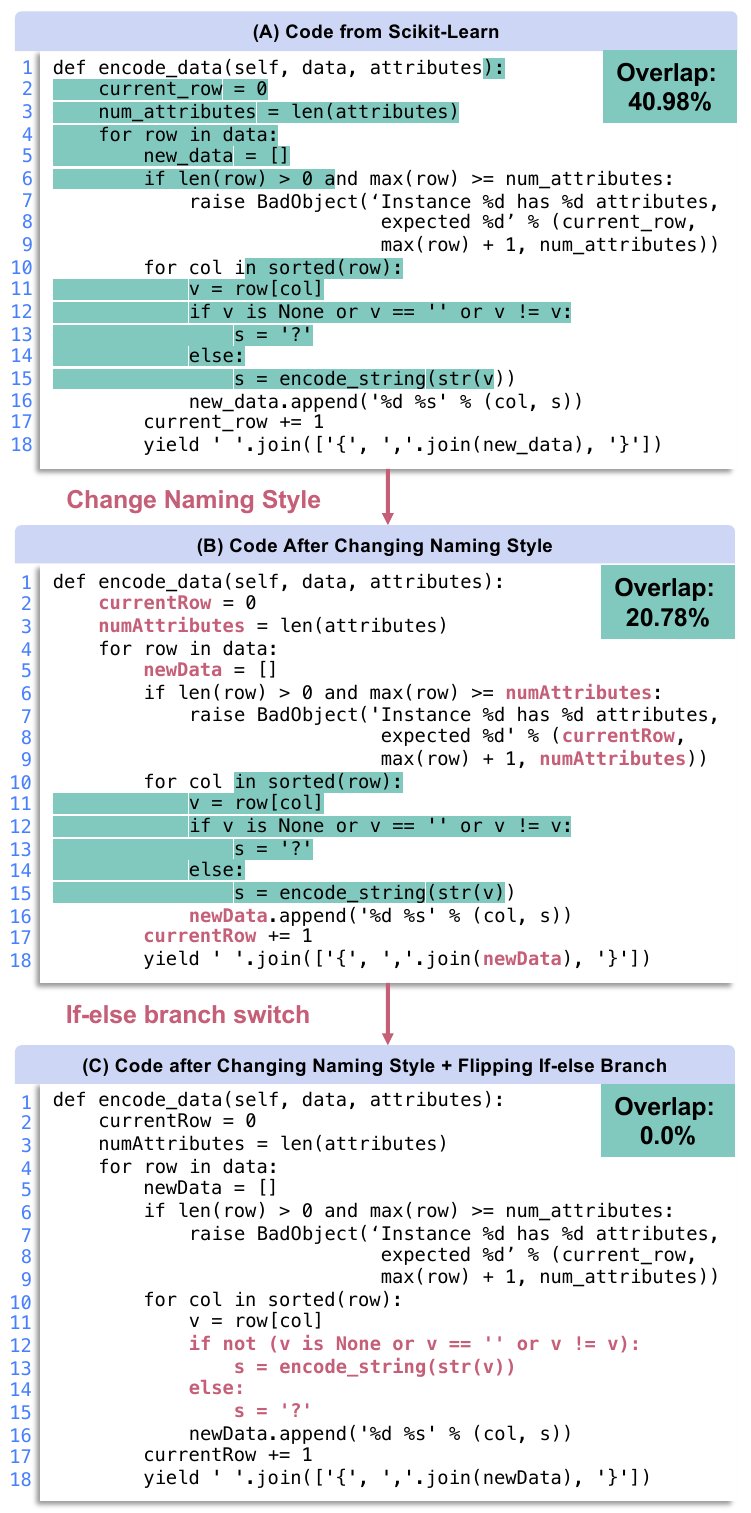}
    \setlength{\abovecaptionskip}{-0pt}
    \setlength{\belowcaptionskip}{-0pt}
    \caption{\textbf{Motivating Example of Code Refactoring Resolving Data Contamination.} \colorbox[HTML]{91C6BE}{Green} highlights the exactly matched characters with the Stack. The result is provided by prior work~\cite{marone2023dataportraits} and can be reproduced online~\cite{url_DataPortraits}.}
    \label{fig:example}
\end{figure} 

Before diving into the design of code refactoring, we use a code snippet from Sklearn~\cite{url_sklearnexample_externals_arff__LODGeneratorData} 
as an example of how code refactoring helps resolve data contamination. The original code is shown in the upper part of Figure~\ref{fig:example}.
To better illustrate the change of code and its overlap with training data, we use \colorbox[HTML]{91C6BE}{green} to highlight the exactly matched characters and use the \textcolor[HTML]{CB6E83}{\textbf{red}} texts to noted the refactored codes. To facilitate the reproduction of this case, we use the overlapping result returned by an online available website~\cite{url_DataPortraits}, 
which is provided by an existing work~\cite{marone2023dataportraits}. Note that the overlapping results are calculated against \textit{the Stack}~\cite{url_stackv1hf}, %\footnote{\texttt{The Stack}: \url{https://huggingface.co/datasets/bigcode/the-stack}}, 
and the whitespace are normalized before checking for overlap.

From Figure~\ref{fig:example}(A), we can see that the initial overlap rate between this code and the training data is 40.98\%, which is high. We first try to disrupt the memorized patterns by changing the naming style from snake case (\ie, connecting words with underscores and using all lower-case letters, \eg, snake\_case) to camel case (\ie, combining words by capitalizing all words following the first word and removing spaces, \eg, camelCase). In the example, three variable names are changed, \ie, \texttt{current\_row}, \texttt{num\_attributes}, and \texttt{new\_data}. After changing the naming style, as shown in Figure~\ref{fig:example}(B), the overlap ratio drops to half, with 20.78\%, resolving the contamination in lines 1-6 in Figure~\ref{fig:example}(A). 

Next, to address the overlaps in lines 10-15, we switch the if-else branches in the code. In particular, we negate the condition in the if-branch and switch the statements in two branches. The resulting codes are shown in lines 15-18 of Figure~\ref{fig:example}(C). After the switch, the overlap successfully drops to 0.0\%.

From this example, we can learn that the key idea of code refactoring in resolving data contamination is to \textit{disrupt the consecutive characters/tokens as much as possible while maintaining the semantic unchanged}. 

\section{Code Refactor Design}\label{sec:refactor-ops}

Bearing the insight observed in Section~\ref{sec:moti}, we then further propose refactoring operators (abbrev. operators) and explore their effectiveness. Ideally, a desired operator should maintain the original code's semantics while perturbing the consecutive tokens as much as possible. 
Also, the operators are better able to be applied automatically without human intervention and assistance, making them easy to use and lightweight. Additionally, the operators are expected to be generalizable and language-agnostic so that they can be generalized to other programming languages. Based on the above three underlines, we then design three categories of refactoring operators, as shown in Figure~\ref{fig:opr}, \ie, \textit{Syntactic Refactoring Operators} (Section~\ref{sec:syn}), \textit{Semantics Refactoring Operators} (Section~\ref{sec:sem}), and \textit{Code Styles Refactoring Operators} (Section~\ref{sec:style}). 

\begin{figure}[tb]
    \centering
    % \vspace{-0ex}
    \includegraphics[width=1.0\linewidth]{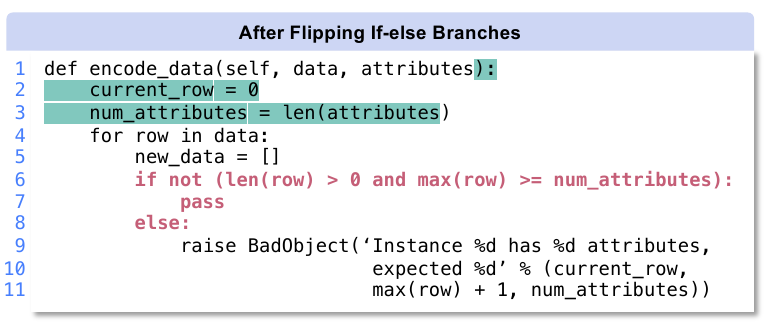}
    \caption{An Example of IF-condition Flipping (\texttt{IFF}) without Else-branch. Original code: Figure~\ref{fig:example}(A).}
    \label{fig:iff}
\end{figure} 

\subsection{{Syntactic Refactoring Operators}}\label{sec:syn}
Operators in this category alter the code syntactic structure while keeping the code semantics untouched. They manipulate the Abstract Syntax Tree (AST) of the code without affecting the original functionalities. In particular, we consider three types of syntactic operators.

\noindent \subsubsection{\ding{202} \textbf{If-condition Flipping  (IFF)}} It is straightforward to negate the if-condition and flip the statements in if- and else-branches. After such a change, the code structures will be changed while the code semantics should not be changed. In particular, if a code snippet contains both if- and else-branches, then \texttt{IFF} can be applied smoothly, as shown in lines 12-15 of Figure~\ref{fig:example} (B) to (C). 
Yet, if a code snippet contains only an if-branch and no else-branch, as shown in lines 6-9 in Figure~\ref{fig:example}(A), then we regard the statement in else-branch as \texttt{pass} and apply the same operation as usual. For the example of lines 6-9 in Figure~\ref{fig:example}(A), the resulting if- and else-branches are shown in Figure~\ref{fig:iff} (lines 6-11).

\noindent \subsubsection{\ding{203} \textbf{Loop Transformation (Loop)}} 
\begin{figure}[t!]
    \centering
    % \vspace{-0ex}
    \includegraphics[width=1.0\linewidth]{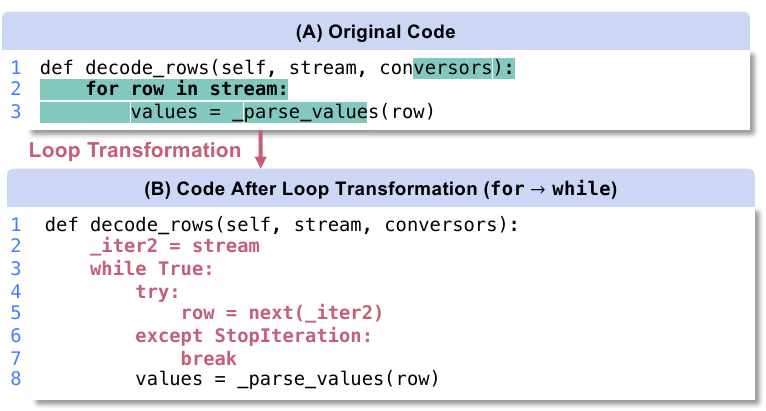}
    \caption{An Example of Loop Transformation (\texttt{Loop})}
    \label{fig:loop}
\end{figure}

Switching \texttt{while} loops with the equivalent \texttt{for} loop also keeps the semantics unchanged. Such transformation is a common semantic-equivalent transformation adopted by prior work~\cite{VJBench23,cao2024concerned}. 
Take \texttt{for} to \texttt{while} loop as an example, shown in Figure~\ref{fig:loop}. First, we identify the loop control variable \texttt{stream} (\ie, over which the loop iterates), and create an iterator \texttt{\_iter2} over it. Note that the name of the iterator will be checked to ensure it does not overwrite the existing variable names in the code snippet.
Second, we formulate the \texttt{while} condition as \texttt{True} and setup the termination condition (\ie, reaching or exceeding the end of the iteration range) to mimic the termination condition in the \texttt{for} loop. To ease the implementation, we use \texttt{try-catch} blocks to iterate over the iterator and terminate till the exception \texttt{StopIteration} is raised. In the \texttt{try} block, we update the control variable using the \texttt{next()} function provided by the iterator. Finally, we transfer the loop body by copying the body of the \texttt{for} loop into the \texttt{while} loop as it is. A similar transformation is applied when transforming \texttt{while} to \texttt{for}. We omit the details due to space limitations.

\noindent \subsubsection{\ding{204} \textbf{Iteration Transformation (Iter)}} 
\begin{figure}[th]
    \centering
    % \vspace{-0ex}
    \includegraphics[width=1.0\linewidth]{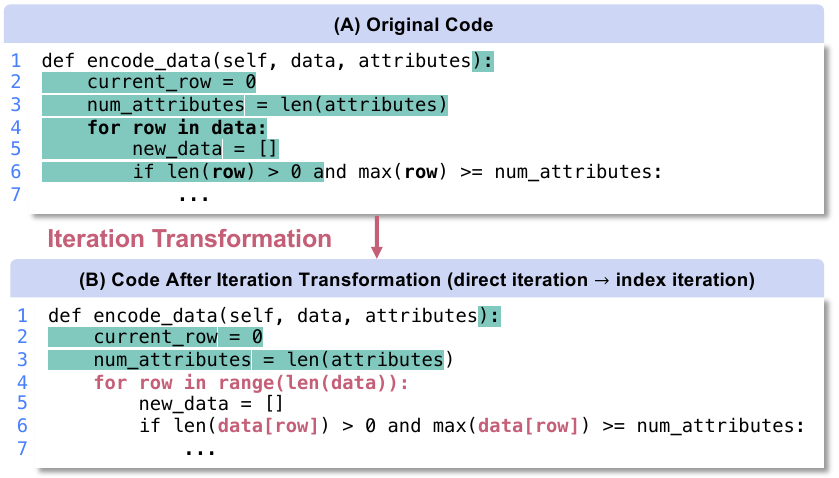}
    \caption{An Example of Iteration Transformation (\texttt{Iter})}
    \label{fig:iter}
\end{figure}

In Python, there are two ways to iterate over an iterable object, \ie, to directly access each element or use indices within a range. The transformation between them keeps the semantics unchanged. An example of such transformation is shown in Figure~\ref{fig:iter}. In particular, line 4 in Figure~\ref{fig:iter}(A) uses direct access to each element in \texttt{data}, denoted by the variable \texttt{row}. After the transformation, the variable \texttt{row} denotes the index within the length range of \texttt{data}. Thus, each iterative element is denoted as \texttt{data[row]}. 

\noindent \subsubsection{\ding{205} \textbf{Commutative Law Flipping (Comm)}} 
\begin{figure}[th]
    \centering
    % \vspace{-0ex}
    \includegraphics[width=1.0\linewidth]{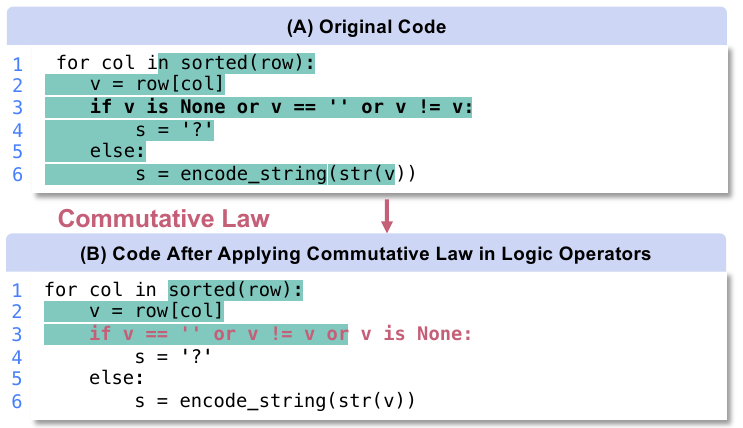}
    \caption{An Example of Commutative Law Shuffling (\texttt{Comm})}
    \label{fig:comm}
\end{figure}

The commutative law in mathematics~\cite{kreuzer2000computational} is a fundamental principle that can be applied to various operators such as logical operations. We implement the commutative law in logical operations $\wedge$ (\ie, \texttt{and} in Python, \texttt{\&\&} in Java) and $\vee$ (\ie, \texttt{or} in Python, \texttt{||} in Java) {{by first grouping operands in each expression according to the operator connecting them, and then shuffling the operands in each group into a new operand sequence}}. 
Figure~\ref{fig:comm} shows an example where the operands \texttt{v is None}, \texttt{v == `'}, and \texttt{v != v} with the same operator \texttt{or} in a logic statement at line 3 are randomly permuted. This reordering of operands (\ie, boolean values and propositions) contributes to disrupting the memorized patterns while preserving the semantics.

\noindent \subsubsection{\ding{206} \textbf{(Class-level) Method Shuffling (Shuf)}}\label{sec:shuf} This operator is designed specifically for class-level code. Shuffling the order of methods within a class will not affect code functionalities and semantics. Such an operator is easy to implement and could effectively disrupt the consecutive tokens in the codes.

\subsection{{Semantic Refactoring Operators}}\label{sec:sem}
Except for the syntactic operators, adding extra semantics or slightly perturbing the original code semantics {without disrupting the function and compatibility of original code} is also applicable. For example, changing identifier names or adding additional context will not change the original code's functionality while introducing slight semantic perturbations into the original code. In particular, according to the changes to the code semantics, we further categorize semantic operators into three subcategories, \ie, \textit{semantic addition} (Sections~\ref{sec:param}, \ref{sec:deco}, \ref{sec:inhr}) and \textit{semantic perturbation} (Section~\ref{sec:renm}). 

\noindent \subsubsection{\ding{207} \textbf{Special Parameter Appending (Param)}}\label{sec:param}
\begin{figure}[th]
    \centering
    % \vspace{-0ex}
    \includegraphics[width=1.0\linewidth]{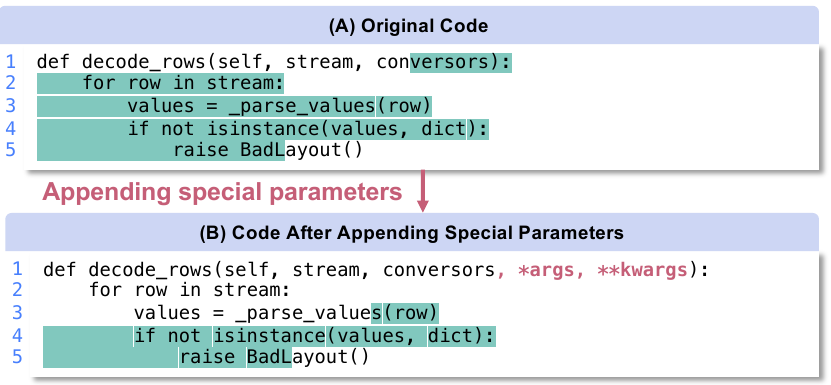}
    \caption{An Example of Special Parameter Appending (\texttt{Param})}
    \label{fig:param}
\end{figure}

A prior work~\cite{cao2024concerned} found out that appending unnamed positional parameters (\texttt{*args}) and keyword parameters (\texttt{**kwargs}) in the parameter list of method declarations tend to alleviate data contamination. Thus, we follow the idea of appending these parameters if they do not exist in the original method declarations. Figure~\ref{fig:param} shows an example where appending special parameters reduces the overlapped tokens. 

\noindent \subsubsection{\ding{208} \textbf{Performance Measurement Decoration (Deco)}}\label{sec:deco} As introduced by a prior work~\cite{cao2024concerned}, adding decorators such as \texttt{@timing} (measuring the execution time) and \texttt{@measure\_memory\_usage} (measuring the memory usage) to Python methods do not change the function's behavior. Thus, we include this operator in our toolkit. 

\noindent \subsubsection{\ding{209} \textbf{(Cross-Class) Inherited method Appending (Inhr)}}\label{sec:inhr} For the classes that inherit from superclasses, copying methods from the superclasses to the (sub)class (if they are not already overridden in the subclass) does not affect the semantics of the subclass. Note that inheritance could be hierarchical, which means that a class may inherit non-overridden methods from its superclass and continue to propagate these methods to its subclasses. After obtaining the inherited methods, an operator of method shuffling (Section~\ref{sec:shuf}) always follows.

\noindent \subsubsection{\ding{210} \textbf{Identifier Renaming (Renm)}}\label{sec:renm} 
\begin{figure}[t!]
    \centering
    % \vspace{-0ex}
    \includegraphics[width=1.0\linewidth]{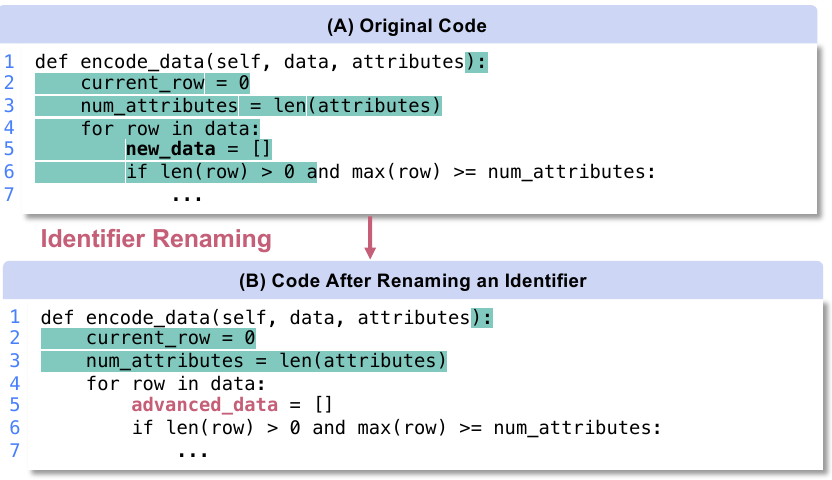}
    \caption{An Example of Identifier Renaming (\texttt{Renm})}
    \label{fig:renm}
\end{figure}

Synonym replacement is a commonly used perturbation in various natural language processing tasks~\cite{SemMT2022,chen2021testing,SIT,chen2021validation,chen2024testingmt}. In the scenario of identifier renaming in code snippets, it posts extra requirements in the words to replace (\ie, the replacement should not be the keywords such as \texttt{while} and \texttt{open}), and the post process of the replacement (\ie, replacing all the occurrence of the replaced identifier). To implement the replacement, we use wordhoard~\cite{wordhoard} as prior work~\cite{cao2024concerned} to replace the identifiers in the source code. Then, we traverse the AST nodes to replace all the occurrences to ensure the code compilation. Figure~\ref{fig:renm} shows an example of identifier renaming. Interestingly, after replacing the variable name \texttt{new\_data} to \texttt{advanced\_data}, the consecutive memorized lines of codes (lines 4-6 in Figure~\ref{fig:renm}(A)) have been disrupted. 

\subsection{{Code Styles Refactoring Operators}}\label{sec:style}

Changing code styles, such as changing naming styles, do not affect the code semantics while interrupting consecutive code characters. Thus, we also include them in our toolkit. 

\noindent \subsubsection{\ding{211} \textbf{Code Normalization (Norm)}}\label{sec:norm} 
\begin{figure}[th]
    \centering
    % \vspace{-0ex}
    \includegraphics[width=1.0\linewidth]{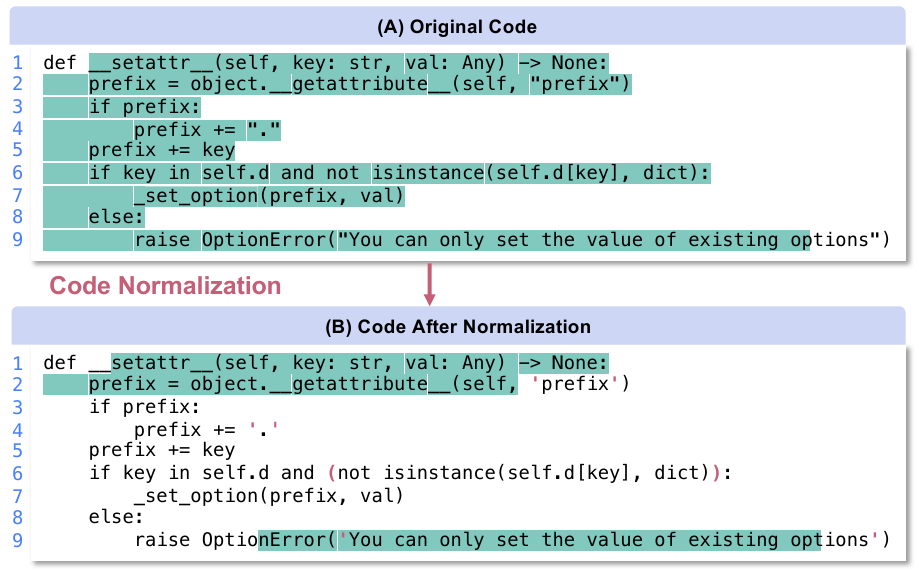}
    \caption{An Example of Code Normalization (\texttt{Norm})}
    \label{fig:norm}
\end{figure}

{Normalizing the code styles, such as unifying single and double quote marks, regularizing the number of spaces, and using parentheses to indicate operation precedence, does not change the integrity of code functionality. Meanwhile, it changes the tokens and thereby helps to interrupt the consecutive token patterns. Figure~\ref{fig:norm} shows an example where the replacement of original double-quotes marks and the addition of parenthesis are applied to the original Python code. These modifications scatter the consecutive tokens, thereby preventing CLMs from recognizing a long token sequence in the memory.}

\textit{2)} \encircled{11} \textit{\textbf{Naming Style Switch (Styl)}:}
The camel case (\eg, camelCase) and lower snake case (\eg, lower\_snake) are two popularly-used naming conventions in programming. Though it is usually the case where Python uses lower snake cases to name variables while Java uses camel cases, the switched situation also allows code compilation and execution. Thus, we flip these two naming styles, once identifying one style, flipping to another. The flipped variable names are applied to all the occurrences. An example is shown in Figure~\ref{fig:example}(A) to (B). After flipping three variables in RFigure~\ref{fig:example}(A), the memorized consecutive lines 2-6 are resolved.

\section{Experiment Preparation}

\subsection{Code Language Models and Data Preparation}

\subsubsection{\textbf{Code Language Models}} We select four representative CLMs widely studied in recent works and used in recent data contamination study~\cite{cao2024concerned}. Table~\ref{tab:model} shows the model information, including the model name, links to the models, training source, and the first release date. In particular, we choose the open-sourced CLMs because their data sources are more transparent compared to commercialized models such as ChatGPT.

\begin{table}[t!]
    \centering
    \renewcommand\arraystretch{1.2}
    \caption{The Studied Code Language Models}\label{tab:model}
    \resizebox{1\linewidth}{!}{
\begin{tabular}{l|l|l|l}
\toprule
Model & Link & Training Source & 1st Release \\
\midrule 
CodeLlama-7b-Instruct & \cite{codellama7b} & GitHub + StackOverflow & Apr, 2023 \\
Starcoder-code-instruct & \cite{starcoderinstruct} & The Stack~\cite{Kocetkov2022TheStack} & May, 2023 \\
StarChat-beta & \cite{starchatbeta} & Github & Jun, 2023 \\
WizardCoder-15B & \cite{wizardcoder} & The Stack~\cite{Kocetkov2022TheStack}, CodeAlpaca-20k & Jun, 2023 \\
\bottomrule
\end{tabular}
}
\end{table}

\begin{figure}[t!]
    \centering
    % \vspace{-0ex}
    \includegraphics[width=1.0\linewidth]{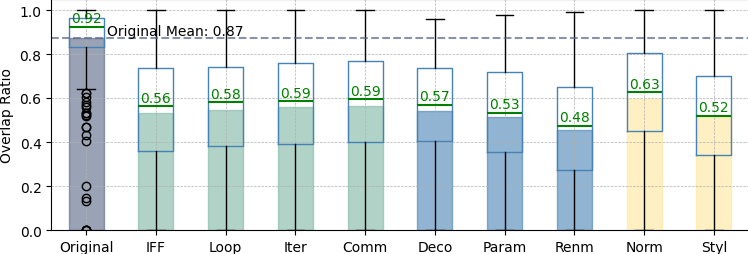}
    \caption{\textbf{RQ1: Data-wise Effectiveness of 9 Method-level Operators in Python Code.} The lower, the better. \colorbox[HTML]{B8D2C8}{Green}/\colorbox[HTML]{97B3CF}{Blue}/\colorbox[HTML]{FCF1C8}{Yellow}:  Syntactical/Semantic/Code style operators. The green texts show the median.}
    \label{fig:overlap1}
\end{figure}

\subsubsection{\textbf{Data Preparation}}\label{sec:data-prepare}
We choose \textit{the~Stack~v1.2}~\cite{Kocetkov2022TheStack} dataset as the code corpus for its representativeness and comprehensiveness, which contains over 2.9 TB data to avoid duplication from {January 1st, 2015} to {March 31st, 2022}. 

\begin{itemize}
    \item \textbf{For RQ1: } 
    To investigate the effectiveness of code refactors in method-level code, we prepare Python code at the method level for refactoring. Specifically, since there are more than ten thousand Python methods every year in \textit{the~Stack~v1.2}~\cite{Kocetkov2022TheStack}, we then follow prior work~\cite{cao2024concerned} to set a 95\% confidence level and 5\% margin of error and sample from the year 2021 (\ie, the last updated date fell in 2021) and only keep the methods with more than three lines of code (LoC) as prior work~\cite{yu2023codereval}. As a result, we sampled 384 Python methods as the original code and then applied all the method-level code refactors to them. These methods include 4 $\sim$ 158 LoC, with an average of 20.68. 
    
    \item \textbf{For RQ2: } To investigate the refactors' effectiveness in class-level and cross-level code, we select three popular Python libraries, \ie, Scikit-learn~\cite{url_sklearn}, Pandas~\cite{url_pandas}, and Numpy~\cite{url_numpy}. We chose them because of their popularity and open access, which give them a higher chance to be included in the training data (evidence can be found in Section~\ref{sec:rq2}). Then, we split these three libraries into Python classes and filter out the classes that {cannot be successfully imported with the standard dependencies of these libraries, \eg, the test classes depending on \textit{pytest}.} 
    In total, there are {2032 classes}. Then, we set a 95\% confidence level and 5\% margin of error and stratified sample from three libraries, resulting in 324 classes {(with an average of 163.55 LoC)} for RQ2, ready for further refactoring.

    \item \textbf{For RQ3:} To study the severity of data contamination in different programming languages, we still sample from \textit{the Stack}~\cite{url_stackv1hf}. We also sampled Java, Rust, and C apart from Python because of their representativeness and popularity, with each 384 method-level code snippet using the same confidence level and margin of error as the sampling strategy. Besides, we sampled data for the years 2018 $\sim$ 2022. In total, there are 4 * 5 * 384 = 7680 code snippets (with an average of 49.73 LoC) to measure the severity of data contamination.

    \item \textbf{For RQ4:} To study the effectiveness of code refactors in Java code, we applied the top-5 effective refactors in Python for Java code. Note that since the minimum function unit of Java is class-level, we sampled 384 Java classes with more than one method with more than 3 LoC as prior work~\cite{yu2023codereval} in the method body so that the method-level refactors are applicable. Then, we compare the effectiveness of the same refactors in different programming languages (\ie, Python and Java), providing potential generalizability of refactors across programming languages. 
\end{itemize}

\begin{figure}[t!]
    \centering
    % \vspace{-0ex}
    \includegraphics[width=1.0\linewidth]{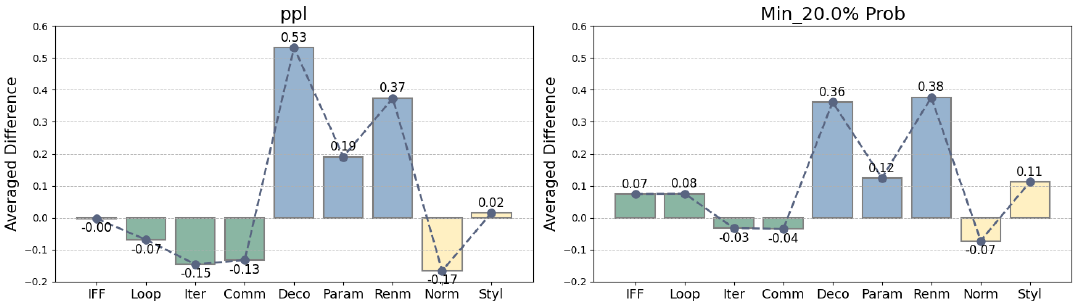}
    \caption{\textbf{RQ1: Two Model-wise Metrics at Method-level Python (model: StarCoder-Instruct)}. Each bar shows the scores of the refactored code subtracted from the scores of the original code. 
    The larger, the better. \colorbox[HTML]{B8D2C8}{Green}/\colorbox[HTML]{97B3CF}{Blue}/\colorbox[HTML]{FCF1C8}{Yellow}:  Syntactical/Semantic/Code style operators. }
    \label{fig:rq1-modelwise}
\end{figure}

\subsection{{Quantification Metrics of Data Contamination}}\label{sec:leakage-metrics}

We measure the severity of data contamination in two folds. First, \textit{\textbf{Data-wise Measurement}} (Section~\ref{sec:overlap}) calculates the overlap with the training sets. Such measurement relies on the full access of the training set~\cite{jiang2024investigating}, and the degree of overlap varies from the training set. However, appearing in the training set does not necessarily mean that the model has fully mastered the content. So we further consider \textbf{\textit{Model-wise Measurement}} (Section~\ref{sec:modelwise}), which measures the model's familiarity with a piece of given content.

\subsubsection{\textbf{Data-wise Measurement}}\label{sec:overlap} We use N-gram matching to quantify the overlap ratio of code snippets against \textit{the Stack}. 

\begin{itemize}
    \item \textbf{N-gram Overlap} (\textit{abbrev}., Overlap) Following prior work~\cite{dodge2021documenting,marone2023dataportraits},
    we quantify the degree of overlap using 50-gram overlap against \textit{the Stack}. 
Specifically, to ease the efforts of data processing, we use the processed data provided by Marone \etal~\cite{marone2023dataportraits}. It is a subset of \textit{The Stack} that is used to train 15.5B StarCoder~\cite{li2023starcoder-paper}, including 800+ GB of code. After processing, a total of 26 GB of documentation artifacts are produced. Then, the overlap is calculated by the number of \textit{overlapped} non-empty characters dividing the total number of non-empty characters. Take Figure~\ref{fig:example} (A) as an example. There are 488 non-empty characters in total, while the number of overlapped non-empty characters (green) is 200, so the overlap ratio is 200/488 = 0.4098.
\end{itemize}

\subsubsection{\textbf{Model-wise Measurement}}\label{sec:modelwise} We consider two popularly used membership inference metrics to measure the familiarity of a model with given code snippets.

\begin{itemize}
    \item {\textbf{Perplexity}} (\textit{abbrev}., ppl) is a measure used to evaluate the predictability of a sequence of words~\cite{jelinek1977perplexity}. Usually, the perplexity measures how ``surprised'' the model is to see a given content~\cite{carlini2019secret}. 
It is defined as the exponentiated average negative log-likelihood of a sequence.

\item \textbf{{MIN-K\% PROB}} was proposed for membership inference attacks~\cite{topkmia2023}. It computes the score using the k\% of tokens with the lowest probabilities. The equation is omitted due to space limitation. 
We set $K$ as $20$ because it produces the most distinguishable differences as shown in prior work~\cite{cao2024concerned}.
\end{itemize}

\section{Evaluation}

We study four RQs listed in Section~\ref{sec:intro}. The data prepared for each RQ can be found in Section~\ref{sec:data-prepare}. The severity of data contamination is quantified using both data-wise and model-wise measurements (see Section~\ref{sec:leakage-metrics}).
The experiments are conducted on a computational infrastructure comprising two NVIDIA RTX 6000 Ada GPUs, each with 48GB VRAM. 

\subsection{RQ1: Effectiveness on Method-level Python Code}

% \begin{table}[t!]
% \centering
% \renewcommand\arraystretch{1.2}
% \caption{Effectiveness of Function-level Operators on Python Code}
% \label{tab:rq1}
% \resizebox{0.9\linewidth}{!}{
% \begin{tabular}{lllll}
% \toprule
%  & Starcoder & Starchat & CodeLlama & WizardCoder \\
%  \midrule
% Param & \cellcolor[HTML]{E4EBF9}0.1966 & \cellcolor[HTML]{F3F6FD}0.1014 & \cellcolor[HTML]{F0F4FC}0.1226 & \cellcolor[HTML]{E4EBF9}0.1986 \\
% Deco & \cellcolor[HTML]{BFD0F1}0.4341 & \cellcolor[HTML]{95B0E7}0.6993 & \cellcolor[HTML]{A1B9EA}0.6239 & \cellcolor[HTML]{BDCEF0}0.4473 \\
% IFF & \cellcolor[HTML]{ECF1FB}0.1464 & \cellcolor[HTML]{F3F6FD}0.1060 & \cellcolor[HTML]{F9FBFE}0.0648 & \cellcolor[HTML]{F1F4FC}0.1189 \\
% Loop & \cellcolor[HTML]{ECF1FB}0.1476 & \cellcolor[HTML]{F2F5FC}0.1106 & \cellcolor[HTML]{F6F8FD}0.0864 & \cellcolor[HTML]{EDF2FB}0.1415 \\
% Renm & \cellcolor[HTML]{BDCEF0}0.4488 & \cellcolor[HTML]{C8D6F3}0.3750 & \cellcolor[HTML]{D2DDF5}0.3149 & \cellcolor[HTML]{BDCEF0}0.4495 \\
% Comm & \cellcolor[HTML]{FDFEFF}0.0367 & \cellcolor[HTML]{FEFFFF}0.0304 & \cellcolor[HTML]{FFFFFF}0.0241 & \cellcolor[HTML]{FDFEFF}0.0389 \\
% Iter & \cellcolor[HTML]{FDFEFF}0.0395 & \cellcolor[HTML]{FFFFFF}0.0275 & \cellcolor[HTML]{FFFFFF}0.0237 & \cellcolor[HTML]{FDFEFF}0.0402 \\
% Styl & \cellcolor[HTML]{E6EDFA}0.1844 & \cellcolor[HTML]{EAEFFB}0.1616 & \cellcolor[HTML]{F3F6FD}0.1041 & \cellcolor[HTML]{E7EDFA}0.1794 \\
% \bottomrule
% \end{tabular}
% }
% \end{table}

\begin{table}[b!]
\centering
\renewcommand\arraystretch{1.1}
\caption{\textbf{RQ1: Model-wise Effectiveness of 9 Operators on Method-level Python Code.} The \colorbox[HTML]{95B0E7}{bluer}, the better. (Metrics: Min\_20\% Prob)}
\label{tab:rq1-model}
\resizebox{1.0\linewidth}{!}{
\begin{tabular}{l|cccc||ccc||cc}
\toprule
 & \multicolumn{4}{c}{Syntactic} & \multicolumn{3}{c}{Semantic} &  \multicolumn{2}{c}{Code Style} \\
 \hline
Models & IFF & Loop & Iter & Comm & Deco & Param & Renm & Norm & Styl \\
 \midrule
 
Starcoder & \cellcolor[HTML]{DCE6F9}0.0740 & \cellcolor[HTML]{DCE6F9}0.0752 & \cellcolor[HTML]{FCDCE3}-0.0329 & \cellcolor[HTML]{FCDAE1}-0.0357 & \cellcolor[HTML]{B2C7EF}0.3617 & \cellcolor[HTML]{CBD9F5}0.1242 & \cellcolor[HTML]{B1C5EE}0.3764 & \cellcolor[HTML]{FABCCB}-0.0724 & \cellcolor[HTML]{D0DBF4}0.1120 \\
Starchat & \cellcolor[HTML]{E7EEFB}0.0511 & \cellcolor[HTML]{E5ECFA}0.0557 & \cellcolor[HTML]{FDE1E7}-0.0273 & \cellcolor[HTML]{FDE4E9}-0.0244 & \cellcolor[HTML]{95B0E7}0.6444 & \cellcolor[HTML]{E9EFFB}0.0465 & \cellcolor[HTML]{B7CAF0}0.3201 & \cellcolor[HTML]{FBCAD5}-0.0549 & \cellcolor[HTML]{CDDBF6}0.1067 \\
CodeLlama & \cellcolor[HTML]{E3EBFA}0.0595 & \cellcolor[HTML]{D9E4F8}0.0811 & \cellcolor[HTML]{F7F9FE}0.0184 & \cellcolor[HTML]{F7F9FE}0.0189 & \cellcolor[HTML]{98B3E8}0.6186 & \cellcolor[HTML]{CBDAF5}0.1174 & \cellcolor[HTML]{B8CBF0}0.3096 & \cellcolor[HTML]{FEF9FA}-0.0052 & \cellcolor[HTML]{D1DDF6}0.0989 \\
WizardCoder & \cellcolor[HTML]{E5ECFA}0.0552 & \cellcolor[HTML]{DBE5F8}0.0778 & \cellcolor[HTML]{FDE5EA}-0.0235 & \cellcolor[HTML]{FDE3E9}-0.0248 & \cellcolor[HTML]{B0C5EE}0.3836 & \cellcolor[HTML]{CAD8F5}0.1349 & \cellcolor[HTML]{B0C5EE}0.3858 & \cellcolor[HTML]{FAC3D0}-0.0637 & \cellcolor[HTML]{CFDAF3}0.1157 \\

\midrule
Average & \cellcolor[HTML]{E3EBFA}0.0600 & \cellcolor[HTML]{DDE7F9}0.0725 & \cellcolor[HTML]{FDEDF0}-0.0163 & \cellcolor[HTML]{FDECF0}-0.0165 & \cellcolor[HTML]{A4BCEB}0.5021 & \cellcolor[HTML]{CDDBF6}0.1057 & \cellcolor[HTML]{B4C8EF}0.3480 & \cellcolor[HTML]{FBCFD9}-0.0491 & \cellcolor[HTML]{CFDAF3} 0.1083

 \\
\bottomrule
\end{tabular}
}
\end{table}

\textbf{\textit{Setup --}} To study the effectiveness of eight method-level operators (shown in Figure~\ref{fig:opr}), we apply each of them to 384 Python methods. In particular, most operators manipulate code over AST, as mentioned in Section~\ref{sec:refactor-ops}, and thereby the code should be syntactically correct and normalized. 
So, except for the operator \textit{Norm}, the effects of other operators are applied to the normalized code. In other words, the effects of other operators can be viewed superimposed on the effect of \textit{Norm}. 

\textbf{\textit{Data-wise Effect --}} The data-wise effectiveness of operators shows in Figure~\ref{fig:overlap1}. It is clear that without refactoring, the overlap of the original code is 92\% (median), which is pretty high. It is reasonable because the original codes are extracted from the training set. After the refactoring, the overlap drops to 48\% $\sim$ 63\%, an average of 35.89\% drop. Among the operators, \textbf{semantic operators witness the largest drop} (average drop: 39.3\%), followed by code style operators (34.5\%) and syntactic operators (34\%). Though the overlap ratio of \textit{Norm} appears the highest among the operators, recall that the other operators are based on the operator \textit{Norm}. Thus, if we look at the reduction brought by each operator independently, the \textit{Norm} operator results in the most significant decrease.

\begin{mdframed}[style=MyFrame]
\textit{\textbf{Finding:}}
Method-level operators can \textbf{\textit{reduce 35.89\% overlap}} with the training set on average. Among them, \textbf{Semantic and code-style operators} are more effective at making the code \textbf{\textit{unlike}} in its form in the training set. 
\end{mdframed}

\textbf{\textit{Model-wise Effect --}} Recap that we consider two model-wise metrics (\ie, perplexity and Min-20\% Prob) to measure the data contamination on specific models, and evaluate against four CLMs (Table~\ref{tab:model}). To better visualize the effects, we subtract the metrics values of all operators from the scores on the original code. A larger difference indicates that the operator makes the model less familiar with the code, thus more effective. Otherwise, a smaller difference indicates less effectiveness.
Figure~\ref{fig:rq1-modelwise} shows the results of two metrics (\ie, ppl and Min\_20\% Prob) for the model StarCoder-Instruct. We can see that the two metrics reflect \textit{similar rankings} to the operators, though the absolute scores vary in two subfigures. Among the operators, \textit{Deco} and \textit{Style} operators are the most effective in making the model less familiar with the code.

Because it is hard to judge the general model-wise effectiveness of operators by only looking at the results on one model, and given the similar rankings offered by two metrics (evident in Figure~\ref{fig:rq1-modelwise}), so we choose Min\_20\% Prob as a representative and show the effectiveness of these operators on the four CLMs together.
As shown in Table~\ref{tab:rq1-model}, we can see that semantic operators, especially \textit{Deco} and \textit{Renm}, are stably making all the studied models unfamiliar with the code, with an average of 0.5021 and 0.3480 increase, thus being the most effective in dealing with model-wise data contamination. Though syntactic operators such as \textit{IFF} and \textit{Loop} can also raise the scores, the increase is negligible. Oppositely, operators \textit{Iter}, \textit{Comm}, and \textit{Norm} make the scores drop slightly, meaning that after these operators, models found more familiar with the code. In other words, syntactic and code-style operators may not stably resolve the data contamination for models.

Such observation echoes prior work~\cite{cao2024concerned}, where they found some syntactic operators (\eg, switching for-loop to while-loop) can sometimes even improve the model performance. We take a further step -- instead of observing the model performance on a specific task, we investigate both model-wise and data-wise contamination, isolating the degree of familiarity with the model and the capability of completing a particular task.

\begin{mdframed}[style=MyFrame]
\textit{\textbf{Finding:}}
Semantic operators \textit{Deco} and \textit{Renm} can stably make all the studied models unfamiliar with the code, showing promising effectiveness in resolving data contamination. Syntactic operators have little effect and sometimes even have a counter-effect. 
\end{mdframed}

\begin{figure}[t!]
    \centering
    % \vspace{-0ex}
    \includegraphics[width=1.0\linewidth]{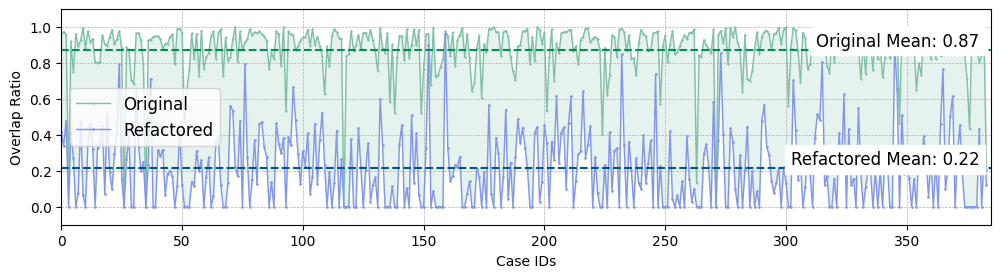}
    \caption{\textbf{RQ1: Best Trial on Method-level Python Code (Data-wise Measurement).}  Green shows the original overlap, while blue shows the overlap after applying all the operators in \name. The lower the overlap with the training set, the less severe the data contamination.
    % Original: Overlap of the original code without refactoring. All: Overlap after applying 9 Method-level operators.
    }
    \label{fig:best-method}
\end{figure}

\textbf{\textit{Best Trial --}} Finally, we apply all operators to see how effectively they can work. Due to space limitations, we only show the data-wise results in Figure~\ref{fig:best-method}. Each point on the X-axis represents the overlap of one of 384 Python method-level code snippets. The green dots represent the overlap ratio between this code snippet and the training set. The blue dots represent the overlap after all the operators are applied to the corresponding code snippet. It is clear that after operators are applied, the overlap drops dramatically, from an average of 87\% to 22\%. Such observation echoes the motivating example shown in Section~\ref{sec:moti}, \ie, after applying two operators, the overlap drops to 0\%, as shown in Figure~\ref{fig:example}.

\begin{mdframed}[style=MyFrame]
\textit{\textbf{Finding:}}\label{subsec:rq1besttryfinding}
The operators we proposed demonstrated exceptional capabilities in addressing data contamination. They successfully reduced the overlap with the training set from 87\% to 22\%, \textbf{achieving a 65\% decrease.}
\end{mdframed}

\subsection{RQ2: Effectiveness on Larger Scale Python Code}\label{sec:rq2}

\textbf{\textit{Setup --}} In view of the effectiveness of operators in method-level Python code, we further explore how they perform on a larger scale. We consider two additional operators, one for class-level and one for cross-class, as shown in Figure~\ref{fig:opr}. In particular, \textit{Inhr} (Cross-class) operator should be applied on more than one class with inheritance relations, and Shuf (Class-level) operator should be applied to a class with more than one method. The detailed design can be found in Section~\ref{sec:refactor-ops}. It is noteworthy that method-level operators are also applicable for larger-scale Python code, so we also evaluate their effectiveness on the data prepared for RQ2 (see Section~\ref{sec:data-prepare} for details). 

\begin{table}[tb]
\centering
\renewcommand\arraystretch{1.3}
\caption{\textbf{RQ2: Model-wise Effectiveness of 11 Operators on Larger Scale Python Code.} The \colorbox[HTML]{95B0E7}{bluer}, the better. (Metrics: Min\_20\% Prob)}
% \caption{Effectiveness of Operators on Cross-Class Python. The \colorbox[HTML]{95B0E7}{bluer}, the better. Oppositely, \colorbox[HTML]{F693AB}{Red} means models are more familiar with the refactored code, so the refactoring operators are less effective.}
\label{tab:rq2}
\resizebox{1.0\linewidth}{!}{
\begin{tabular}{l||ccccc||cccc||cc}
\toprule
 & \multicolumn{5}{c||}{Syntactic} & \multicolumn{4}{c||}{Semantic} & \multicolumn{2}{c}{Code Style} \\
 \hline
Models & IFF & Loop & Iter & Comm & Shuf* & Deco & Param & Inhr* & Renm & Norm & Styl \\
 \midrule
 
Starcoder & \cellcolor[HTML]{EDF2FB}0.056 & \cellcolor[HTML]{FEF9FA}-0.006 & \cellcolor[HTML]{FBD9E0}-0.037 & \cellcolor[HTML]{FCDEE4}-0.032 & \cellcolor[HTML]{D0DCF6}0.063 & \cellcolor[HTML]{9AB4E9}0.306 & \cellcolor[HTML]{FCDAE1}-0.036 & \cellcolor[HTML]{FBC2CF}-0.246 & \cellcolor[HTML]{9AB4E9}0.306 & \cellcolor[HTML]{FBD9E0}-0.044 & \cellcolor[HTML]{F3F7FD}0.016 \\
Starchat & \cellcolor[HTML]{FCFDFF}0.011 & \cellcolor[HTML]{FDEFF2}-0.015 & \cellcolor[HTML]{FCDAE1}-0.036 & \cellcolor[HTML]{FCE5EA}-0.025 & \cellcolor[HTML]{DCE6F8}0.046 & \cellcolor[HTML]{9EB7EA}0.287 & \cellcolor[HTML]{FBD4DC}-0.090 & \cellcolor[HTML]{FBBDCB}-0.315 & \cellcolor[HTML]{A4BBEB}0.259 & \cellcolor[HTML]{FBD9E0}-0.040 & \cellcolor[HTML]{F6F8FD}0.013 \\
CodeLlama & \cellcolor[HTML]{F1F5FC}0.046 & \cellcolor[HTML]{F2F6FD}0.017 & \cellcolor[HTML]{FEFEFF}0.006 & \cellcolor[HTML]{FBFCFF}0.013 & \cellcolor[HTML]{D0DCF6}0.063 & \cellcolor[HTML]{AEC3EE}0.208 & \cellcolor[HTML]{FEF4F6}-0.010 & \cellcolor[HTML]{FBC5D1}-0.223 & \cellcolor[HTML]{A8BFEC}0.236 & \cellcolor[HTML]{FEFFFF}0.004 & \cellcolor[HTML]{E3EBFA}0.037 \\
WizardCoder & \cellcolor[HTML]{F0F4FC}0.049 & \cellcolor[HTML]{FAFBFE}0.007 & \cellcolor[HTML]{FCE1E6}-0.029 & \cellcolor[HTML]{FCE6EA}-0.025 & \cellcolor[HTML]{C9D7F4}0.076 & \cellcolor[HTML]{95B0E7}0.328 & \cellcolor[HTML]{FBD9E0}-0.038 & \cellcolor[HTML]{FBC0CD}-0.266 & \cellcolor[HTML]{96B1E8}0.325 & \cellcolor[HTML]{FCDAE1}-0.037 & \cellcolor[HTML]{F0F4FC}0.021 \\
\midrule
Average & \cellcolor[HTML]{F2F6FD}0.040 & \cellcolor[HTML]{FFFFFF}0.001 & \cellcolor[HTML]{FEF4F6}-0.024 & \cellcolor[HTML]{FEF7F9}-0.017 & \cellcolor[HTML]{D0DDF6}0.062 & \cellcolor[HTML]{9FB8EA}0.282 & \cellcolor[HTML]{FDEBF0}-0.044 & \cellcolor[HTML]{F58BA5}-0.263 & \cellcolor[HTML]{9FB8EA}0.281 & \cellcolor[HTML]{FDF2F5}-0.029 & \cellcolor[HTML]{EFF3FC}0.022

 \\
\bottomrule
\end{tabular}
}
\end{table}

\begin{figure}[th]
    \centering
    % \vspace{-0ex}
    \includegraphics[width=1.0\linewidth]{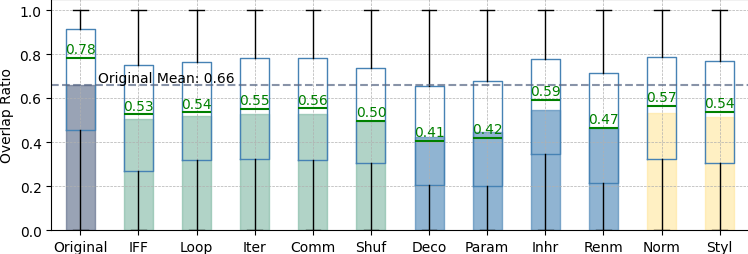}
    \caption{\textbf{RQ2: Data-wise Effectiveness of 11 Operators in larger-scale Python Code.} The lower, the better. \colorbox[HTML]{B8D2C8}{Green}/\colorbox[HTML]{97B3CF}{Blue}/\colorbox[HTML]{FCF1C8}{Yellow}:  Syntactical/Semantic/Code style operators. The green texts show the median.}
    \label{fig:overlap-class}
\end{figure}

\textbf{\textit{Data-wise Effect -- }} The result is visualized in Figure~\ref{fig:overlap-class}. We can see the original overlap is up to 78\% on median.  After 11 operators are applied, the decrease in overlap is obvious, with 37\% (78\% - 41\%) drop at best. Also, the rankings are similar as that on method-level Python code (as shown in Figure~\ref{fig:overlap1}, where semantic operators also achieve the largest drop. It shows the generalizability of method-level operators on larger-scale Python code.

\begin{mdframed}[style=MyFrame]
\textit{\textbf{Finding:}}
The 11 operators show promising results \textbf{\textit{on larger-scale Python code, reducing a maximum of 37\% overlap rate}} with the training set. The method-level operators are \textbf{\textit{generalizable}} to larger scales.
\end{mdframed}

Additionally, zoom in on the two additional operators, \textit{Shuf} and \textit{Inhr}, we can see that shuffling the methods in classes (\textit{Shuf}) does help to decrease the overlap, from 78\% to 50\%, with a 28\% drop. Yet, appending inherited methods from the superclass raises the overlap slightly on average. This is because although inserting inherited code may interrupt the continuity of the original code, the newly inserted code may also introduce additional overlap, so the overall overlap increases. 

\begin{mdframed}[style=MyFrame]
\textit{\textbf{Finding:}}
\textbf{\textit{Class-level \textit{Shuf} can effectively reduce the overlap}} with the training set by 28\%. Although the cross-class \textit{Inhr} can work in some cases, overall, the final overlap ratio increases slightly due to the introduction of duplicate code.
\end{mdframed}

\textbf{\textit{Model-wise Effect -- }} Similar as RQ1, we show the Min\_20\% Prob scores in four models in Table~\ref{tab:rq2}. Also, because the scale of Python code goes large, we 
{append no more than three inherited methods to avoid lengthy classes (\textit{Inhr}) and rename no more than three variables due to the time cost to request the online synonym library (\textit{Renm}).} in a code snippet. From Table~\ref{tab:rq2}, we can see that there are six operators that are still effective on larger Python codes (\ie, positive average values, shown in blue), achieving at most 0.282 increment. 
However, it is evident that the operators struggle more on larger scale code than on the method-level code, with fewer blocks showing blue (\ie, effective). \textit{Deco} and \textit{Renm} still play the most dramatic role, with a 0.281 increase in Min\_20\% Prob scores, followed by the class-level operator \textit{Shuf}. Method-level syntactic operators (\ie, \textit{IFF}, \textit{Loop}, \textit{Iter} and \textit{Comm}) have little help in resolving data contamination on larger-scale Python code, which is not surprising given the observation in RQ1. 

It is noteworthy that \textit{Param} can take effects on method-level Python code (see Table~\ref{tab:rq1-model}) while it does not work in class-level code. This may be because the special parameters \texttt{*args} and \texttt{**kwargs} frequently appear in larger-scale Python code while rarely appearing in single methods. So, adding special parameters in method-level code can interrupt models' memorization patterns while ineffective for larger-scale code.

\begin{figure}[th]
    \centering
    % \vspace{-0ex}
    \includegraphics[width=1.0\linewidth]{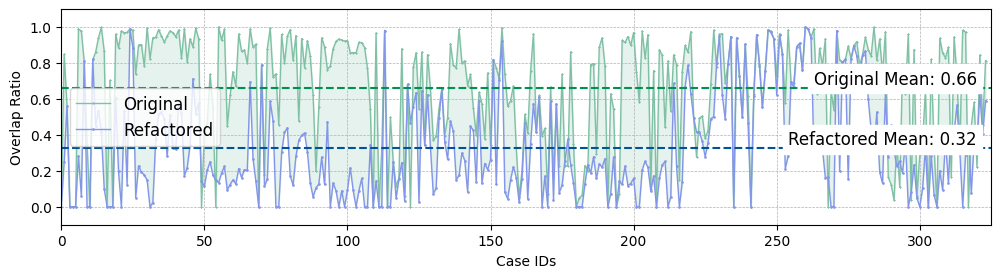}
    \caption{\textbf{RQ2: Best Trial on Class-level Python Code.} Green shows the original overlap, while blue shows the overlap after applying all the operators in \name once. The lower the overlap, the less severe the data contamination.
    % Original: Overlap of the original code without refactoring. All: Overlap after applying 11 Class-level operators.
    }
    \label{fig:best-class}
\end{figure} 

\begin{mdframed}[style=MyFrame]
\textit{\textbf{Finding:}}
In terms of the model-wise data contamination measurement, \textbf{\textit{most operators are still effective}} on larger-scale Python code, although the degree of effectiveness is significantly less than on smaller-scale code. \textit{Deco} and \textit{Renm} still play the most dramatic role, while \textit{Param} is on longer effective on larger-scale Python code. 

\end{mdframed}

\textbf{\textit{Best Trial --}} We also investigate the optimal effectiveness of applying all operators on the class-level subjects and show the data-wise results in Figure~\ref{fig:best-class}, where each point on the X-axis represents one of the 324 Python class-level code snippets. 
The overlap is found to drop dramatically after all operators are applied (blue dots) compared to the original value (green dots), from an average of 66\% to 32\%. This observation is consistent with the one on method-level cases. It further confirms our last finding in Section~\ref{subsec:rq1besttryfinding} and echoes the motivating example.

\subsection{RQ3. Contamination In Other Programming Languages}

\textbf{\textit{Setup --}} The previous two RQs studied the severity of data contamination in Python at method- or larger level. In this RQ, we further explore the contamination in different programming languages and how it changes over time. We focus on Java, C, and Rust due to their popularity and show data-wise overlapping from the year 2018 to 2022.

\begin{figure}[t!]
    \centering
    % \vspace{-0ex}
    \includegraphics[width=1.0\linewidth]{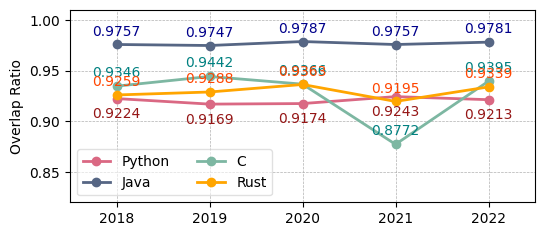}
    \caption{RQ3: Contamination Severity of Various Programming Languages Changing Over Time}
    \label{fig:year}
\end{figure}

The data-wise results are shown in Table~\ref{fig:year}. It is clear that Java overlaps with the training set the most, reaching an average of 98\%. Python, C, and Rust have similar overlap ratios, averaging from 88\% to 94\%. Meanwhile, the trend of overlap of each programming language has been relatively stable over time, except for C, which had a significant drop in 2021. The model-wise results show similar trends to the data-wise results, but we omit the figures due to space limitations.

\begin{mdframed}[style=MyFrame]
\textit{\textbf{Finding:}}
The severity of data contamination is similar across different programming languages, with Java having the highest at 98\% on average, and other programming languages close to 94\%. Moreover, from 2018 to 2022, data contamination has remained relatively stable each year. 
\end{mdframed}
\subsection{RQ4. Effectiveness of Operators in Java Code}

\textbf{\textit{Setup --}} Finally, we study the possibility of generalizing the operators to other programming languages. In particular, Deco, Param, and Iter are not applicable to Java, while Comm and Styl have little effect, so we implement \textit{IFF}, \textit{Loop}, \textit{Renm}, and \textit{Norm} for Java, and apply them in the 384 Java classes (the preparation can be found in Section~\ref{sec:data-prepare}). 

\begin{figure}[tb]
    \centering
    % \vspace{-0ex}
    \includegraphics[width=1.0\linewidth]{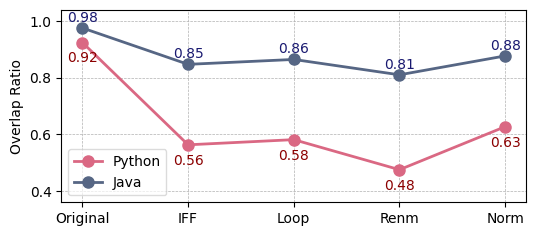}
    \caption{RQ4: Comparison of Operators on Java and Python}
    \label{fig:pyjavacompare}
\end{figure}

\begin{table}[b!]
\centering
\renewcommand\arraystretch{1.2}
\caption{\textbf{RQ4: Model-wise Effectiveness of 4 Operators on Method-level Java Code.} The \colorbox[HTML]{95B0E7}{bluer}, the better. (Metrics: Min\_20\% Prob)}
\label{tab:rq4}
\resizebox{1.0\linewidth}{!}{
\begin{tabular}{l||cc||c||c}
\toprule
 & \multicolumn{2}{c||}{Syntactic} & \multicolumn{1}{c||}{Semantic} & Style \\
 \hline
Models  & IFF & Loop & Renm & Norm \\ 
 \midrule
 
Starcoder & \cellcolor[HTML]{B3C7EE}0.08839778 & \cellcolor[HTML]{F2F6FD}0.0314298 & \cellcolor[HTML]{D0DDF5}0.08340778 & \cellcolor[HTML]{F9C1CE}-0.0264679 \\
CodeLlama & \cellcolor[HTML]{E1EAF9}0.07257575 & \cellcolor[HTML]{ECF2FB}0.04642795 & \cellcolor[HTML]{AFC4EE}0.08905231 & \cellcolor[HTML]{F9FBFE}0.01626699 \\
WizardCoder & \cellcolor[HTML]{DEE8F9}0.07884058 & \cellcolor[HTML]{F3F6FD}0.03070538 & \cellcolor[HTML]{9FB8EA}0.09190729 & \cellcolor[HTML]{FAD3DC}-0.018788 \\
Starchat & \cellcolor[HTML]{95B0E7}0.09348985 & \cellcolor[HTML]{F0F5FC}0.03712159 & \cellcolor[HTML]{C1D2F2}0.08593003 & \cellcolor[HTML]{FBDEE4}-0.0143738 \\

\midrule
Average & \cellcolor[HTML]{D0DEF5}0.08332599 & \cellcolor[HTML]{F0F5FC}0.03642118 & \cellcolor[HTML]{B8CBF0}0.08757435 & \cellcolor[HTML]{FCE6EA}-0.0108407

 \\
\bottomrule
\end{tabular}
}
\end{table}

\textbf{\textit{Data-wise Effect --}} 
The effectiveness of Java operators is shown in Figure~\ref{fig:pyjavacompare} (blue). The initial overlap is around 98\%, which is pretty high. After refactoring, at most 17\% (= 0.98 - 0.81) drop can be achieved by \textit{Renm}.

\textbf{\textit{Model-wise Effect --}} 
The model-wise results of four operators on Java are shown in Table~\ref{tab:rq4}. Semantic \textit{Renm} and syntactic \textit{IFF} are similarly effective on Java among four operators, reaching 0.08+ improvement. Yet, the degree of such improvement is minor, calling for more effective operators on Java.

\begin{mdframed}[style=MyFrame]
\textit{\textbf{Finding:}}
Most migrated operators show a positive effect on resolving Java code contamination, which shows the generalizability of operators in \name. 
While considering the improvement is subtle, how to effectively refactor Java is still an open question.
\end{mdframed}

\section{Related Work}\label{sec:related}

\subsection{Data Contamination in Large Language Models}
Several studies~\cite{sainz-etal-2023-nlp-contam,dataContamination2023,taskContamination2023,balloccu-etal-2024-leak,tirumala2022memorization,kandpal2022deduplicating,schick2020s,magar2022data} have explored the data contamination threat in large language models (LLMs). 
Brown \etal~\cite{extracting21} analyzed the contamination in GPT-3, showing that the performance between contaminated and clean data is not necessarily different. 
Carlini \etal~\cite{tirumala2022memorization} found that the memorization of LLMs grows with model size, training data duplicates, and prompt length. 
Kandpal~\etal~\cite{kandpal2022deduplicating} identified that the success of privacy attacks on LLMs could be attributed to the sequence duplication in the training set. 
Razeghi~\etal~\cite{schick2020s} studied the correlations between LLMs' performance and the frequency of terms and observed that LLMs perform better on inputs with more frequent terms.  
Magar~\etal~\cite{magar2022data} pre-trained and fine-tuned models (\ie, BERT) from scratch, with a controlled training set, to identify factors that affect model exploitation.
Balloccu~\etal~\cite{balloccu-etal-2024-leak} studied the impact of the indirect data contamination, \ie, using user feedback as the training source. 
These works studied the severity of data contamination, while our study investigated the effectiveness of code refactoring operators in alleviating data contamination.

\subsection{Detection of Data Contamination}

As data contamination becomes inevitable, methods have been proposed to detect it. One of the most commonly used methods is n-gram matching~\cite{nips20fewshot,chowdhery2023palm,touvron2023llama,marone2023dataportraits}, which partitions text into sequences of N consecutive characters/tokens and compares comparing these sequences to find patterns or similarities with the training data. For example, Marone~\etal~\cite{marone2023dataportraits} offers a website~\cite{url_DataPortraits} %~\footnote{DataPortraits: \url{https://dataportraits.org/}}
to compare with three data sources, including \textit{The pile}~\cite{url_pile}, %\footnote{\texttt{The pile}: \url{https://pile.eleuther.ai/}}, 
\textit{The Stack}~\cite{url_stackv1hf}, %~\footnote{\texttt{The Stack-V1}: \url{https://huggingface.co/datasets/bigcode/the-stack}}
and \textit{The Stack-V2}~\cite{url_stackv2hf}. %~\footnote{\texttt{The Stack-V2}: \url{https://huggingface.co/datasets/bigcode/the-stack-v2}}. 
Recently, Deng \etal~\cite{deng2023investigating} proposed a mask-and-fill method to detect contamination in multiple-choice questions and raised concerns about data contamination in Question-Answering benchmarks 
such as TruthfulQA~\cite{lin2021truthfulqa} and MMLU~\cite{hendrycks2020measuring}. 

Another related task, {{Membership inference attacks}} (MIAs), determines whether a given data is contained in the model's training data and can be used as a hint for data contamination. 
Various metrics are proposed to infer data membership, including LOSS~\cite{loss-mia}, reference models~\cite{extracting21}, perplexity~\cite{jelinek1977perplexity}, Zlib Entropy~\cite{gailly2004zlib},
Neighborhood attack~\cite{mattern2023membership}, Min-k\% Prob~\cite{topkmia2023},~\etc
Though MIAs are extensively studied in traditional deep learning models, 
the research on MIA in LLMs is limited. Duan~\etal~\cite{duan2024membership} conducted a large-scale MIA on LLMs and found that MIAs barely outperform random guessing across varying LLM sizes and domains. Recently, Dong~\etal~\cite{generalization2024} detected data contamination by identifying the peakedness of LLM's output distribution and showed potential.

Our work differs from these works in purposes, \ie, they detect the existence of data contamination while we introduce effective code refactors that address data contamination.

\subsection{Resolution of Data Contamination}

Various benchmarks~\cite{jimenez2024swebench,mbpp2021,humaneval,evalplus,yu2023codereval,du2023classeval,VJBench23} have been proposed to avoid data contamination via involving code after LLMs' cut-off-date. 
Jain \etal~\cite{jain2024livecodebench} traced and recorded the latest code contests from LeetCode, AtCoder, and CodeForces, aiming at providing a contamination-free code contests benchmark for LLMs evaluation.

Instead of proposing new benchmarks, another bunch of works explores code refactoring.
For example, Wu~\etal~\cite{VJBench23} adopted identifier renaming and code structure change to the code they collected. Besides, several studies~\cite{fan2023large,poldrack2023ai,noever2023chatbots} leverage code assistants such as GPT-4 to do the code refactoring. 

Our work differs from these works in two ways. First, we avoid using AIs to eliminate reintroducing variants in our study. Second, their refactoring heavily relies on manual crafting, while our refactors are conducted automatically. Third, these works did not compare the change in the severity of data contamination after the refactoring.

\section{Threats to Validity}

We acknowledge several threats to the validity of our conclusions. 
First, training set selection bias. Our study only uses \textit{the Stack} as the training set, while there are other training sets for code, such as the \textit{CodeSearchNet}~\cite{husain2019codesearchnet} and \textit{CodeParrot}~\cite{hendrycksapps2021}. We chose \textit{the Stack} because it is popularly used in CLMs training (more than 60+ CLMs trained on it~\cite{url_stackclms}, and %~\footnote{\url{https://huggingface.co/models?dataset=dataset:bigcode/the-stack}}), 
has a long time span ({January 1st, 2015} to {March 31st, 2022}).
Second, the sampling may not be representative. For each RQ, we sampled 324 $\sim$ 384 Python/Java code at the method/class level to experiment.
The sampling may introduce bias, so the conclusion may vary. To alleviate this threat, we carefully set up the confidence level as 95\% and the margin of error as 5\% as the sampling scale. We also filter out the low-quality code (\eg, too short or extra long) to ensure the quality of sampled data. 
Third, the semantic operators we implemented may inadvertently change the code semantics, making the model find the code semantics strange and thus exacerbating model-wise data contamination.

\section{Conclusion}\label{sec:conclusion}
In this paper, we introduce \name, an off-the-shelf automated refactoring toolkit for Python and Java {to mitigate data contamination in CLM evaluation}. We perform quantitative analysis of the change in data contamination severity after applying different operators and identify effective operators under various settings (\ie, model-wise or dataset-wise, method-level, or larger-scale). We make \name online available to facilitate the studies towards more lightweight and effective CLM data contamination solutions. It enables software companies to more accurately assess the true performance of CLM-based techniques before their adoption.

\clearpage

\balance
\bibliographystyle{IEEEtran}
\bibliography{Tex/reference}

\end{document}